# Infrared nanospectroscopic mapping of DNA molecules on mica surface

*Irma Custovic, Nicolas Pocholle, Eric Bourillot, Eric Lesniewska, Olivier Piétrement\**

Laboratoire Interdisciplinaire Carnot de Bourgogne (ICB), UMR CNRS 6303, Université Bourgogne Franche-Comté, 9 Avenue Alain Savary, 21078 Dijon Cedex, France

ABSTRACT

Significant efforts have been done in last two decades to develop nanoscale spectroscopy techniques owning to their great potential for single-molecule structural detection and in addition, to resolve open questions in heterogeneous biological systems, such as protein-DNA complexes. Applying AFM-IR technique has become a powerful leverage for obtaining simultaneous absorption spectra with a nanoscale spatial resolution for studied proteins, however the IR-AFM investigation of DNA molecules on surface, as a benchmark for a nucleoprotein complexes nanocharacterization, has remained elusive. Herein, we demonstrate methodological approach for acquisition of IR-AFM mapping modalities with corresponding absorption spectra based on two different DNA deposition protocols on spermidine and $Ni^{2+}$ pretreated mica surface. The nanoscale IR absorbance of distinctly formed DNA morphologies on mica are demonstrated through series of IR-AFM absorption maps with corresponding IR spectrum. Our results thus demonstrate the sensitivity of IR-AFM nanospectroscopy for a nucleic acid research with an open potential to be employed in further investigation of nucleoprotein complexes.

TEXT

The assessment of structure-depended function and interactions of biomolecules has been greatly hinged on the development of diverse spectroscopic techniques[1–4]. In that context, a particularly applicable approach is vibrational spectroscopy, owning to its leverage to retrieve bond-specific chemical information assigned to molecular characterization and its correlative structural properties[5–9]. The most essential cellular biological processes depend on complex interplay between nucleic acids and proteins and a quantitative assessment of these interactions is crucial for understanding mechanisms which govern DNA replication, transcription, recombination or DNA repair[10]. While the latter are increasingly well-characterized by conventional vibrational spectroscopy on ensemble level[11–15], however understanding the nanoscopic nucleoprotein-level interactions, or lack thereof, remains a prominent challenge. In general, low-dimensional biological phenomena occur at nanometer scale size, whereas the applications of conventional vibrational spectroscopy are constrained due to low spatial-resolution performance set by optical diffraction limit, which could at its best, approach 1 μm depending on the excitation source. Since bulk spectroscopy could be challenging to apply for biological heterogeneous systems, such us DNA-protein complexes due to acquisition of the chemical information averaged on the ensemble level of molecular spices, the behaviour of individuals and their internalization mechanism has been addressed by single-molecule spectroscopy which could provide a profound impact on how we understand bimolecular processes[16,17].

A widely adapted label-based technique for single-molecule detection in biochemical research is fluorescence spectroscopy providing a temporal fluorescence-encoded resolution of biomolecules under studies[18,19]. In particular, the fluorescence spectroscopy has been effectively used for investigation of distinct structural behaviour with an access to molecular observable dynamic resulting from protein-DNA interactions[20,21]. Along with label-based assessment, the understanding of interactions among biomolecules could be contingent on the label-free spatial imaging modalities, with high sensitivity at nanoscale area over the nanometer length scale of relevance to biological systems.

Apropos of various chemical nanoimaging techniques, two nanoscale analogues of Raman and IR spectroscopy coupled with Atomic Force Microscopy, TERS and AFM-IR techniques facilitates a common platform enabling nanometer spatially resolved chemical spectroscopy without substantial sacrifice of sensitivity and spatial resolution[22–27]. Both techniques became of particular interest for biological systems, as has been reported recently to demonstrate nanoscale (<20 nm) vibrational insight in constitutes of bacteria[28], individual vesicles[29], oligomer and fibrillar aggregates during amyloid formations[30], protein aggregates[30–33], protein-based process[34,35], towards descending size cellular constructs such as methylation status in single human metaphase chromosome[36] as well secondary structure of single protein molecule[37].

Since the advent of atomic force microscopy (AFM), and among its essential role in biology processes, deoxyribonucleic acid (DNA) is considered as a benchmark sample, against which new technical developments are tested [38,39]. In current state-of-the-art, TERS technique has provided tremendous sensitivity and applicability in nucleic acid research[23,27], all the more pursuing the crucial demands for a sequencing procedure of single stranded DNA molecule and chemically identifying a single base-pair resolution[40]. As complemental nanospectroscopy method sharing the same scanning probe platform, yet based on different physical mechanism, IR-AFM advantageously combines the nanoscale spatial resolution provided by AFM together with the chemical information offered by IR spectroscopy[22,24]. Unlike conventional IR spectroscopy, AFM cantilever is used as detector to sense localized IR-absorption induced thermal expansion thus providing nanoscale resolution-related chemical information through IR spectrum[41]. The outstanding advantage of IR-AFM technique is thus direct measuring IR absorption as a function of position across the sample creating colorized AFM-IR chemical maps that correspond to distribution of chemical species localized at the nanoscale. Although recently, Knowls et al. have demonstrated first acquisition of infrared absorption spectra and chemical maps of protein at the single molecule level[37], surprisingly to our knowledge, IR-AFM has not been used for nanocharacterization of a single DNA molecule, to date.

The recent improvement of AFM-IR sensitivity by integration of resonance enhanced cantilever oscillation with high repetition rate laser allows measurement of ultra-thin samples down to monolayer, thus providing an interest to be employed into DNA research, despite its small dimeter (< 2 nm) and demanding surface properties[42,43]. We propose here to reach this challenging goal by nanochemical mapping of DNA molecules deposited onto mica surface thus pushing the horizons further in applicability of IR-AFM technique for nucleic acid research. Intrinsically, the thermo-mechanical properties of sample and the dynamics of AFM cantilevers are crucial to acquire AFM-IR data, and as such, has a great potential to simplify methods of nanoscale characterization of DNA molecules. In that context, DNA plasmids and mica surface, as underlying substrate, could be used as a standard sample to perform AFM-IR nanocharacterization. Given the reported knowledge of DNA deposition on surfaces for conventional AFM studies, one can state that atomically flat mica surface is preferable substrate that permits a broad range of DNA deposition methods while providing high SNR images[44]. Therefore, we present first reported methodological approach for acquisition of IR-AFM absorption mapping modality and corresponding spectrum of DNA on mica surface employing short-term deposition based on biochemical relevant protocols.

Primarily, summarizing the reported knowledge, two main strategies are established to bind negatively charged DNA onto negatively charged mica surface: 1.) polyelectrolyte-coated mica surface with polyamines-based chemical compounds such as spermidine, spermine or poly-L-lysine. [45,46] and 2.) through metal counterions mediated DNA adsorption[47,48]. From the first DNA adsorption strategy, we have chosen spermidine-pretreatment method which usually forms self-assembled monolayer with exposed positively charged amino groups $NH^{3+}$ that results in attraction of DNA through Coulomb interaction. The second chosen strategy for DNA binding upon mica surface is based on mica pretreatment by transition metal cations such as $Ni^{2+}$ and $Zn^{2+}$, which simultaneously enhances the DNA fixation and reduces the repulsive contribution, proposed by model of electrical double-layer force[47].

Following the deposition onto mica surface, IR-AFM characterization of DNA requires specific conditions to be conductive including DNA installation upon IR-AFM prism. Prior to installation upon $CaF_2$ prism, the DNA-mica sample was cleaved on samples' backside by scotch tape in order to obtained as transparent-possible mica surface for efficient IR laser light transmissions (see Figure S4 in ESI). Based on our empirical evidence, yet reproducible one (see Figure S4 in ESI) it is important to note that thickness of mica installed upon prism effects the amplitude but not the contrast of DNA on recorded AFM-IR absorption map.

Since the AFM tip–sample contact dynamics critically determine the AFM-IR outcomes, the contemplated choice of AFM-IR parameter's values such as scan speed, set-point and IR laser tune specifications are crucial for obtaining high quality AFM-IR data of DNA.

The infrared nanocharacterization of DNA molecules was conducted using AFM-IR instrument (nanoIR ™, Anasys Instruments) as described in general details of Experimental Section of Electronic Supplementary Information. The laser power is mostly set to 15 % and adjusted to obtain a clean and distinct 'IR-hotspot' in the center of the image which corresponds to the location where the AFM tip is in contact with the sample. The IR-AFM imaging of DNA samples is conducted in contact mode at a scan rate of 0.6 Hz with a narrow-window of a low setpoint values (-2.5 V until 0.25 V) avoiding the destruction contact-mode influence of the tip upon sample. As factor of environment conditions could promote the alternation of DNA contrast on IR-AFM absorption map (see Figure S6 a) and b) in ESI) we propose to set temperature at 25 °C and below with humidity less than 25%.

**Nanoscale chemical imaging of DNA network formed upon spermidine pretreated mica surface**

Biogenic polyamines, such as spermine and spermidine, are small organic polycations involved in diverse DNA-based biological processes and show prominent ability to bind and condensate single DNA and chromatin[45,49,50]. In addition polyamines are essential to AFM spreading for AFM imaging in all condition, and especially at high salt concentration.[51] Typical AFM image of DNA deposited upon spermidine-functionalized mica surface with corresponding profile section is presented in Figure S1. The porous-like extended DNA network observed in Figure S1 with its profile section of approximately 1.6 nm suggests that DNA molecules are attracted to each other, yet not strong enough to generate fully spherical-condensed structure. As a first attempt towards IR-AFM analysis of DNA molecules, we have deliberately chosen DNA network formed upon spermidine pretreated mica surface, due to its extended DNA morphology which could provide enhanced collection of AFM-IR signal as an outcome. The second cantilever oscillation mode was chosen for optimizing the cantilever ringdown signal at its frequency centre of 171 kHz using a frequency window of 50 kHz. The infrared laser focus was optimized at 1728 cm$^{-1}$, 1654 cm$^{-1}$, 1648 cm$^{-1}$, and 1550 cm$^{-1}$(see details below Figure S3). Figure 1 presents AFM-IR topography images (a-d) recorded simultaneously with corresponding (e-h) AFM-IR absorption maps based on intensity of the IR signal at selected wavenumber values, annotated above the bottom row of Figure 1 (e-h). Each set (AFM image and AFM-IR map) were recorded directly after previous image once the optimization of IR laser light is done.

The absorption IR-AFM maps recorded at 1728 cm$^{-1}$, 1654 cm$^{-1}$ and 1646 are taken as referenced from FTIR spectrum showing nanoscale IR absorption of DNA highest at 1728 cm$^{-1}$ regarding to underlying mica surface (Figure 1e), while gradually attenuated at 1654 m$^{-1}$ and 1646 cm$^{-1}$ (Figure 1f and g) while the lowest IR absorption contrast occurred at 1550 cm$^{-1}$.

The IR-AFM spectra are collected in the range between 1520 cm$^{-1}$ and 1820 cm$^{-1}$ where peaks principally correspond to in-plane double-bond stretching vibrations characteristic for various nucleotide base pairing schemes in DNA[52] (see Figure 2).

**Nanoscale chemical imaging of single molecule DNA on mica surface**

The second strategy for binding DNA onto mica surface is based on Ni$^{2+}$ preincubated mica and three-steps based protocol with narrow ionic conditions[53] (see ESI Experimental section). The AFM image of 2.3 kbp single DNA molecule and corresponding contour length are presented in Figure S2. The quantification of the single DNA molecules' contour length agreed with 20% reduction in contour length while using $MgCl_2$ based buffer, which is consistent with the DNA adopting mixed A-form and B-form conformation[54,55]. For IR-AFM nanocharacterization of single DNA molecule deposited onto mica surface, we have chosen second cantilever oscillation mode for optimizing the cantilever ringdown signal at its frequency center of 171 kHz using otherwise reduced frequency window of 25 kHz, instead of 50 kHz utilized for AFM-IR nanocharacterization of DNA network. The infrared laser focus was priory optimized at 1728 $cm^{-1}$, referenced from the highest DNA absorbance on IR-AFM absorption map in Figure 1e.

Figure 3 presents AFM-IR imaging of single DNA molecule on mica surface recorded with infrared laser focus optimized at 1633 $cm^{-1}$ (see details below Figure S5 in ESI). The higher IR absorbance is contributed to single DNA molecule regarding to underlying mica surface which is monitored by AFM-IR absorption map (Figure 3b) as bright-red DNA contour. The acquired and averaged spectra (Figure 3c) based on two positions on DNA molecule (outline by blue crosses in Figure 3a and 3b) obtained broadening of peaks of amide bands with absorption maxima at 1618 $cm^{-1}$. Besides, the prominent spectral band has been found in region between 1520-1558 $cm^{-1}$ with embedded peaks at 1544 cm-1 assigned to Amide II band[55,56]. Additionally, band feature between 1640-1670 $cm^{-1}$ is assigned to carbonyl stretching modes $\upsilon(C=O)$ of DNA residues overlapping with amide I[6,52]. Within this band, two peaks are notable at 1656 $cm^{-1}$ and 1664$cm^{-1}$, assigned to amide I and $C_2=O_2$ strength of cytosine, respectively, along with shoulder peaks at 1652 $cm^{-1}$ and 1672 $cm^{-1}$. A band with lower intensity is identified at 1568 which corresponds guanine ring vibration doublet[57]. Spectral feature observed at 1589 $cm^{-1}$ is attributed to C-C phenyl ring stretchs.

Figure 4 represents evident shifting of single DNA molecule absorbance once the pulse laser tune is fixed at 1664 cm$^{-1}$ and 1550 cm$^{-1}$, illustrating the sensitivity of IR detection.

**Discussion**

Consecutive registration of DNA network absorbance through AFM-IR maps at selected wavenumbers 1728 cm$^{-1}$, 1654 cm$^{-1}$ and 1648 cm$^{-1}$ represent good agreement with obtained FT-IR data. In fact, the highest absorbance of DNA network is mapped at optimized wavenumber 1728 cm$^{-1}$ (outline by bright red porous contour in Figure 1e). The discrepancy between the highest signal amplitude at FTIR spectrum (1648 cm$^{-1}$) and the highest absorbance at AFM-IR maps (1728 cm$^{-1}$) could possibly originate from different area's size of signal collection. However, these founding were preliminary evidence for the DNA network nanoscale screened absorbance respond based on series of AFM-IR absorption maps. The absence of a mica pretreated signal contribution to the reported AFM-IR spectra of DNA was also determined in Figure S7 in ESI. The influence of DNA sample preparation on IR-AFM spectrum have not been discussed to date. The interaction of DNA with complexes is mainly due to bases. Hence, the evidence of DNA-spermidine interaction should be observed in 1550-1800 range assigned to in-plane vibrational stretching. The prominent band appearing in range between 1700 cm$^{-1}$ and 1770 cm$^{-1}$ with a peak at 1739 cm$^{-1}$ in Figure 2 b could possibly embed C=O stretching overlapping with a signal of spermidine-DNA interaction[56,58,59]. It is reported that a major shift of absorbance intensity was mainly observed for a guanine band at 1718 in polyamine-DNA complexes in major groove within this range. Since the polyamines may fulfil a role in gene regulation through major groove, we show that our results obtained by AFM-IR could become the inception to for future study of polyamine-DNA induced structural effects at nanoscales. In addition, the acquired AFM-IR spectrum of single DNA molecule consists protruding contribution of nucleobases vibrational stretching. The consecutive registration of single DNA absorption through AFM-IR maps (see Figure 5) at selected wavenumbers 1634 cm$^{-1}$, 1664 cm$^{-1}$ and 1674 cm$^{-1}$ represent dominative absorbance fingerprint of dT, dC, dA residues and absorbance of poly(dG-dC). poly(dG-dC) for 1634, 1664 and 1674, respectively[52,60,61]. Therefore, based on acquired absorption maps and spectrum with a resolution of single DNA molecule, IR-AFM could become a powerful strategy for spectroscopic nanocharacterization of nucleoprotein complexes.

**Conclusion**

We show in this work the first AFM-IR analysis of DNA, organized in network or on single molecule. The IR signal is highly sensitive to experimental conditions, such as temperature or humidity, but with right settings, we demonstrate that it is possible to chemically map DNA sample with a nanometer scale and with a high sensitivity. Indeed, to prepare DNA network we condensated DNA with spermidine, which bind strongly to DNA, and we detected the presence of spermidine on IR spectrum. We believe that achievement of high AFM-IR sensitivity of deposited DNA molecules on mica surface are prerequisite for IR-AFM nanospetroscopic characterization of DNA-protein complexes in future work leading for instance to the identification of protein in muti-components assembly.

FIGURES

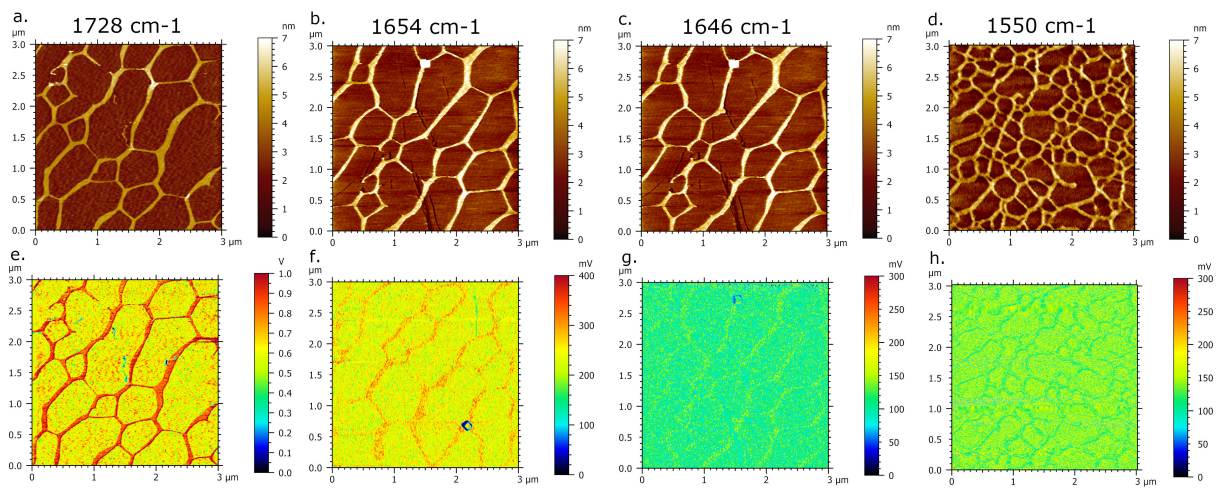

**Figure 1.** AFM-IR imaging of DNA network deposited onto spermidine-functionalized mica surface. a)-d) AFM morphology map and corresponding e)-h) IR absorption maps of DNA network with a highest contrast-absorption of DNA acquired at 1728 cm$^{-1}$, gradually attenuated at 1654 and 1646 cm$^{-1}$ and highest contrast-absorption of background at 1550 cm$^{-1}$.

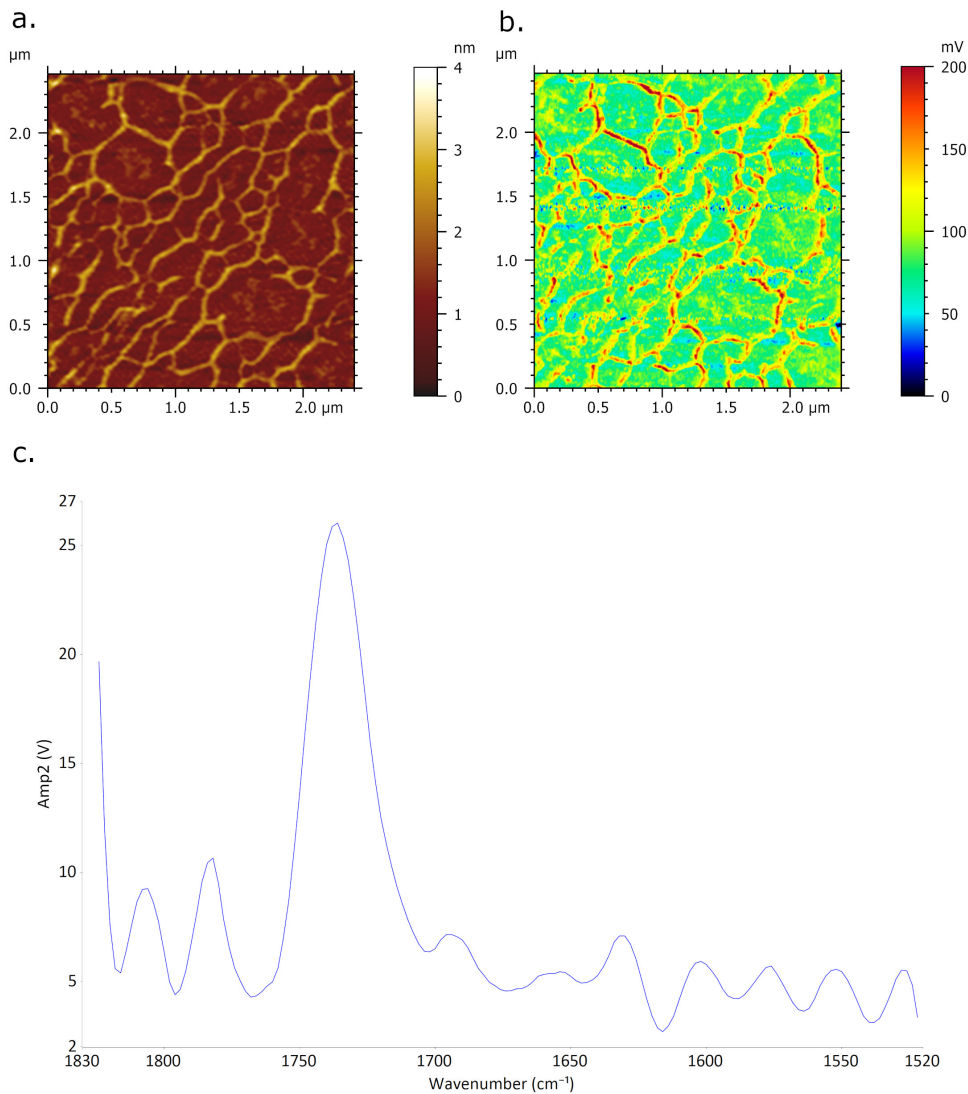

**Figure 2.** AFM-IR imaging of DNA network formed upon spermidine pretreated mica surface. Size of imaged area: 1.0 x 1.0 µm. a) AFM topography image of DNA and corresponding b) AFM-IR absorption map of DNA network recorded at optimized wavenumber 1728 cm$^{-1}$ showing DNA of higher absorbance (bright red porous contour) regarding to underlying mica surface. c) AFM-IR spectra acquired and collected from two positions of DNA (blue cross in Figure 3a and 3b) showing prominent broaden band peaks of C=O stretching with absorption maxima at 1737 cm$^{-1}$.

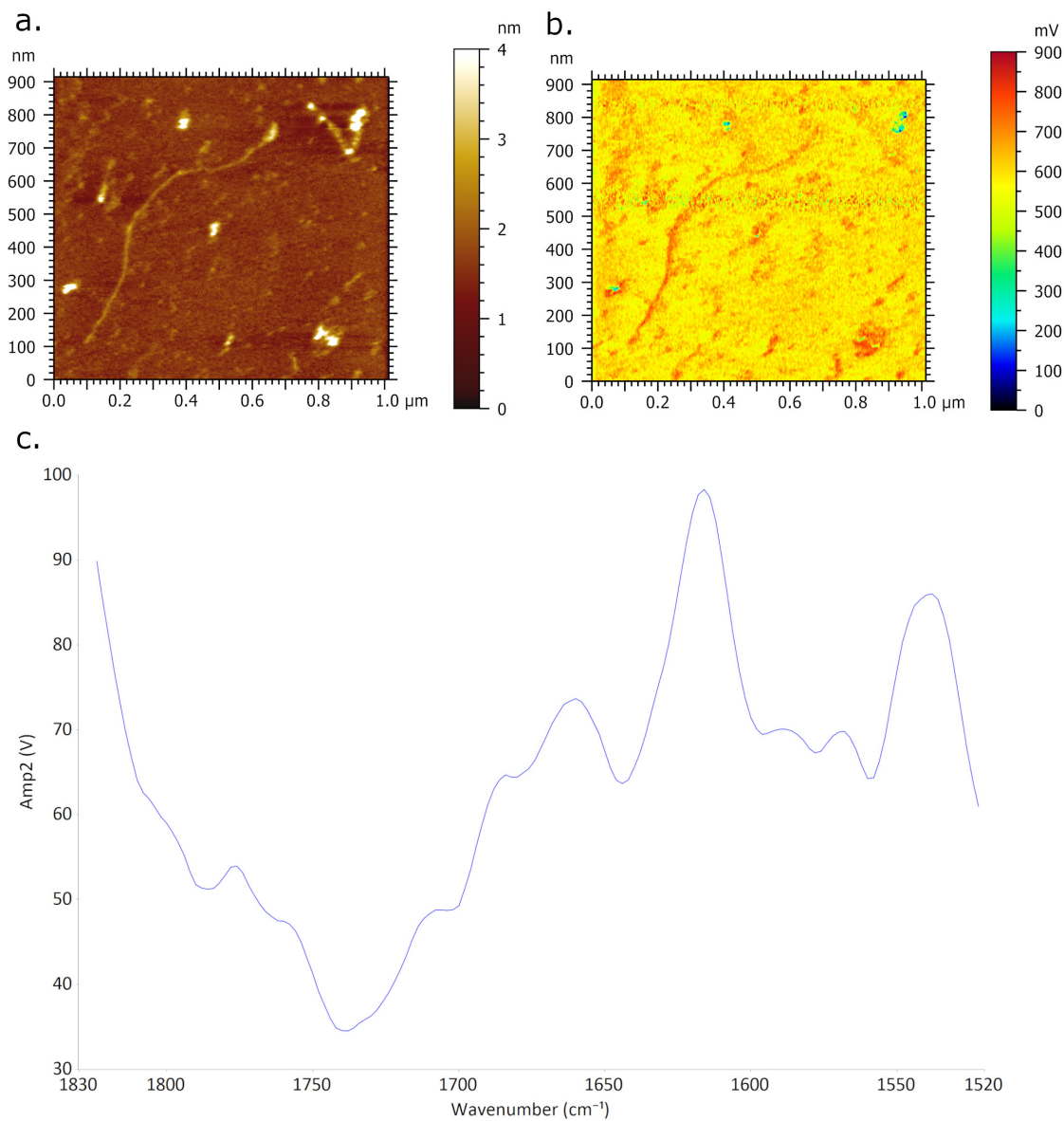

**Figure 3**. AFM-IR imaging of single DNA molecule deposited upon Ni$^{2+}$ pretreated mica surface. Size of imaged area: 1.0 x 1.0 μm. a) AFM topography image of DNA and corresponding b) AFM-IR absorption map of single DNA recorded at optimized wavenumber 1633 cm$^{-1}$ showing DNA of higher absorbance (bright red contour) regarding to mica surface. c) AFM-IR spectra acquired and collected from two positions of DNA (two blue crosses in Figure 3a and 3b) showing prominent broadening peaks of amide bands with absorption maxima at 1618 cm$^{-1}$.

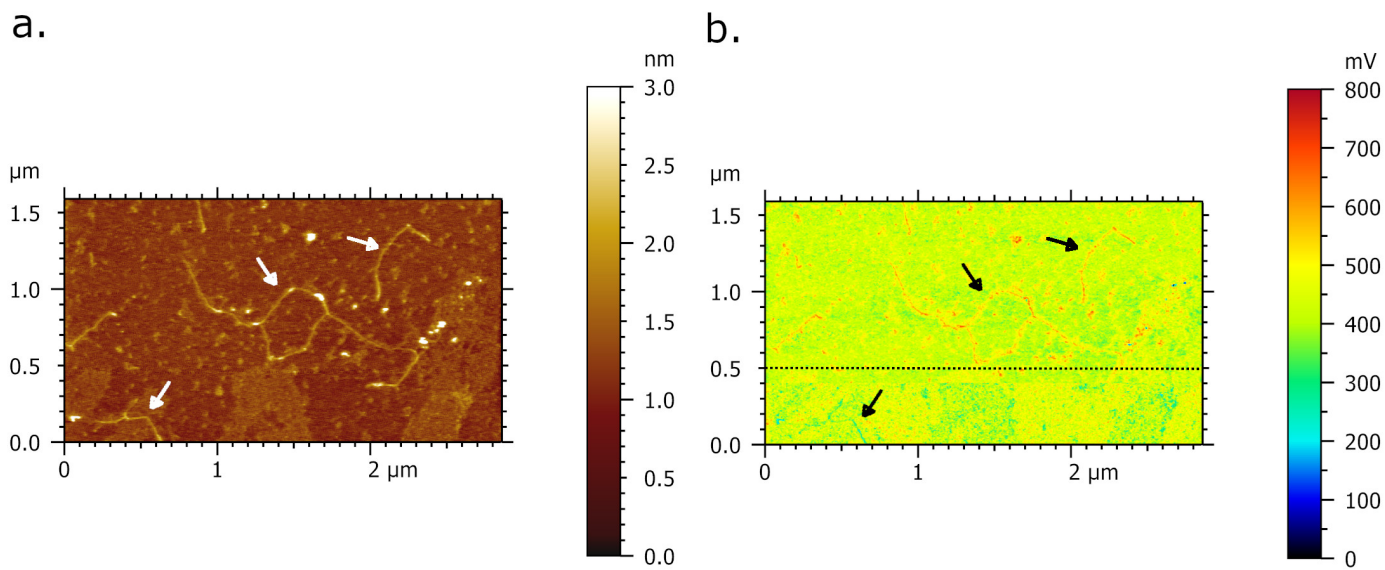

**Figure 4** AFM-IR imaging of single DNA molecule deposited onto $Ni^{2+}$-functionalized mica surface. a) AFM morphology map and corresponding b) AFM-IR absorption maps of single DNA with a higher (red contour) and lower (blue/green contour) absorbance acquired at laser pulse tune configuration with fixed frequencies at 1654 $cm^{-1}$ (above dashed line) and 1550 $cm^{-1}$ (below dashed line), respectively.

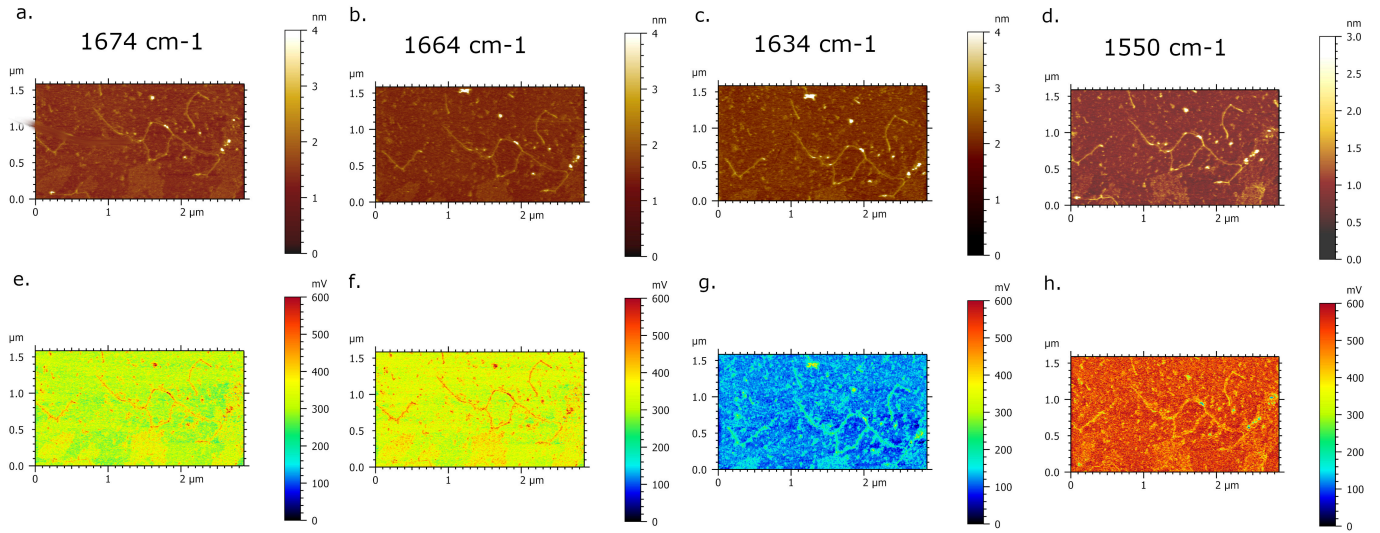

**Figure 5** AFM-IR imaging of DNA single molecule deposited onto nickel functionalized mica surface. a)-d) AFM morphology map and corresponding e)-h) IR absorption maps of DNA single molecule with a highest contrast-absorption of DNA acquired at 1674 cm$^{-1}$, 1664 cm$^{-1}$, 1634 cm$^{-1}$ and 1550 cm$^{-1}$, respectively.

## ASSOCIATED CONTENT

**Supporting Information**.

The Supporting Information is available free of charge at

## AUTHOR INFORMATION

**Corresponding Author**

Olivier Piétrement - Laboratoire Interdisciplinaire Carnot de Bourgogne (ICB), UMR CNRS 6303, Université Bourgogne Franche-Comté, 9 Avenue Alain Savary, 21078 Dijon Cedex, France ; https://orcid.org/ 0000-0002-0018-7202 ; olivier.pietrement@u-bourgogne.fr

**Author Contributions**

The manuscript was written through contributions of all authors. All authors have given approval to the final version of the manuscript.

**Funding Sources**

OP acknowledges funding from the CNRS Mission pour l'Interdisciplinarité (MI-DynAFM-DNARep 2018_273085) and région Bourgogne-Franche-Comté (AAP Région 2020 - ANER - Projet AFMdynDNA). This work has been also supported by the EIPHI Graduate School (contract ANR-17-EURE-0002).

## ACKNOWLEDGMENT

The authors would like to acknowledge Dr. C.-H. Brachais and M.-L. Léonard for their assistance and advices during FTIR experiments.

## NOTES

The authors declare no competing financial interest

P53 with Poly(ADP-Ribose) and DNA Induce Distinct Changes in Protein Structure as Revealed by ATR-FTIR Spectroscopy. *Nucleic Acids Res.* **2019**, *47* (9), 4843–4858. https://doi.org/10.1093/nar/gkz175.

(13) Tse, E. C. M.; Zwang, T. J.; Barton, J. K. The Oxidation State of [4Fe4S] Clusters Modulates the DNA-Binding Affinity of DNA Repair Proteins. *J. Am. Chem. Soc.* **2017**, *139* (36), 12784–12792. https://doi.org/10.1021/jacs.7b07230.

(14) Ruggeri, F. S.; Šneideris, T.; Vendruscolo, M.; Knowles, T. P. J. J. Atomic Force Microscopy for Single Molecule Characterisation of Protein Aggregation. *Arch. Biochem. Biophys.* **2019**, *664* (1), 134–148. https://doi.org/10.1016/j.abb.2019.02.001.

(15) Zhang, Y.; Iwata, T.; Yamamoto, J.; Hitomi, K.; Iwai, S.; Todo, T.; Getzoff, E. D.; Kandori, H. FTIR Study of Light-Dependent Activation and DNA Repair Processes of (6 À 4) Photolyase. *Biochemistry* **2011**, *50* (18), 3591–3598. https://doi.org/10.1021/bi1019397.

(16) Moerner, W. E.; Kador, L. Optical Detection and Spectroscopy of Single Molecules in a Solid. *Phys. Rev. Lett.* **1989**, *62* (21), 2535–2538. https://doi.org/10.1103/PhysRevLett.62.2535.

(17) Moerner, W. E.; Shechtman, Y.; Wang, Q. Single-Molecule Spectroscopy and Imaging over the Decades. *Faraday Discuss.* **2015**, *184*, 9–36. https://doi.org/10.1039/C5FD00149H.

(18) Weiss, S. Measuring Conformational Dynamics of Biomolecules by Single Molecule Fluorescence Spectroscopy. *Nat. Struct. Biol.* **2000**, *7* (9), 724–729.
20

*Electronic Supporting Information*

Infrared Nanospectroscopic mapping of DNA molecule on mica surface

Irma Custovic, Nicolas Pocholle, Eric Bourillot, Eric Lesniewska, Olivier Pietrement*

Laboratoire Interdisciplinaire Carnot de Bourgogne (ICB), UMR CNRS 6303, Université Bourgogne Franche-Comté, 9 Avenue Alain Savary, 21078 Dijon Cedex, France

**EXPERIMENTAL SECTION**

**Sample preparation for AFM experiments**

We purchased puC19 DNA plasmid (New England Biolabs), and we linearized it with EcoRI enzyme (New England Biolabs). After 60 minutes of incubation, we rinsed the solution with Tris-HCl 10 mM (pH 7.5), NaCl 1M buffer solution through Amicon Ultra-0.5 30 kD (Millipore). DNA is finally eluted to 10 µg/ml in TE buffer [10mM Tris-HCl (pH 7.45), 1mM EDTA] and aliquots of 50 µL were frozen and stored at -20°C.

1. **DNA network deposition on mica surface**

We diluted DNA in ionic buffer [10 mM Tris, 20 mM $NiCl_2$ and 5 mM $MgCl_2$] down to 5 µg/mL and flash frozen. We purchased Spermidine 0.1 M solution (Sigma-Aldrich Chemicals Co) stored at 4°C and diluted at 50 µM. On freshly cleaved mica surface (Muscovite Mica Sheets 15 mm diameter disk, V1 quality, Electron Microscopy Science) we deposited 5 µL drop of spermidine and incubated for 1-2 minutes. Afterwards, we added 5 µL of DNA ionic-buffer solution upon spermidine-pretreated mica area and incubated for 3 minutes. Each deposition step (spermidine and DNA-buffer solution) was followed by drying mica surface with gentle touch of filter paper.

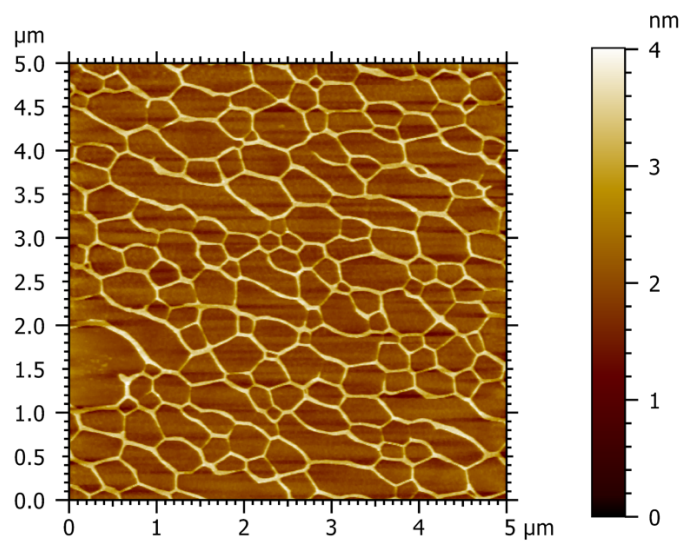

**Figure S1**. AFM image of DNA formed network on spermidine pretreated-mica surface.

2. **Single molecule DNA deposition on mica surface**

We diluted DNA in ionic-deposition buffer [10 mM Tris-HCl (pH=7.45), 10 mM MgCl2, 25 MM KCl] down to 5 µg/mL and stored at 4 °C overnight. The protocol of single molecule DNA deposition is adapter from Perkins et al. We pretreated mica surface with a 20 µL drop of 100 mM $NiCl_2$ (Sigma-Aldrich) onto the freshly cleaved mica for 1 min followed by rinsing with 50 mL of ultrapure water. The mica was then quickly dried by touching filter paper and completely drying the surface. Immediately after drying mica surface, we deposited 20 µL of DNA in ionic-deposition buffer where the concentration of the DNA was 1.5 µg/mL. After 2 s, we gently rinsed the surface with ~1 mL of deposition buffer followed by additional 8 mL of dewetting-based rinsing in portions of 150 µL. Finally, the surface was gently rinsed with 2 mL of imaging buffer [10 mM Tris (pH 7.5), 10 mM $NiCl_2$ +25 mM KCl]. During deposition and rinsing, solutions were kept at room temperature (19 °C for the room containing our AFM). At all other times, the salt reagents were kept at 4 °C. Buffers were filtrated with filter paper of 0.2 µm porosity each day from concentrated stocks.

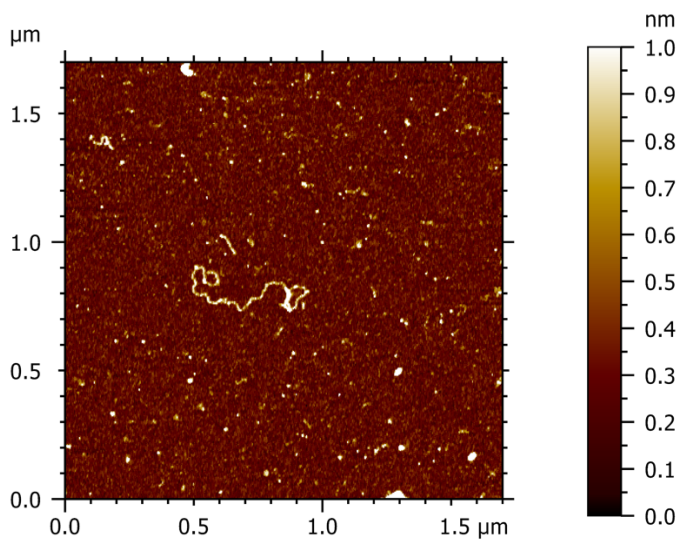

**Figure S.2** Single molecule DNA deposited on Ni-pretreated mica surface

**FT-IR Experiment**

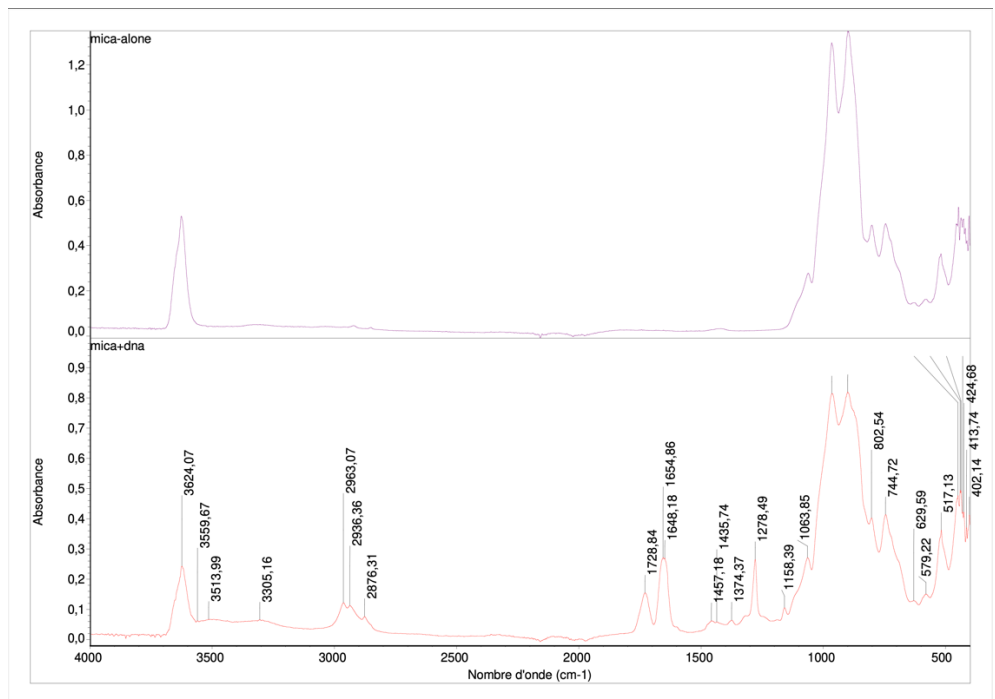

**Figure S3**. FT-IR spectrum of a) bear mica surface b) spermidine-pretreated mica surface with deposited DNA network

Prior to IR-AFM imaging, we employed FTIR spectroscopy, stored in *Electronic Supplementary Information*, to examine chemical-vibrational respond of deliberately chosen DNA network sample, due to extended DNA morphology which could provide enhanced IR signal as an outcome. For the spectrum of DNA deposited on spermidine-pretreated mica surface, characteristic IR absorption peaks can be observed at 1728 cm$^{-1}$, 1654 cm$^{-1}$, 1646 cm$^{-1}$ in domain between 1800 cm$^{-1}$ and 1500 cm$^{-1}$. In this domain, the DNA peaks are principally corresponding to in-plane-double bond stretching vibrations of the bases which are characteristic of various base pairing schemes.

**Nano-IR AFM spectroscopy measurements**

The DNA sample on mica surface was exfoliated from a reverse side by scotch tape until very thin and transparent mica surface is obtained, (see Figure S4) in order to efficiently enhance IR light transmission. After, the DNA sample is transferred upon $CaF_2$ pyramid prism (source Crystran) and fixed with type N immersion liquid (Leica Microsystems). We performed nanospectroscopy measurements combined with AFM by NanoIR1 system (Anasys Instruments Inc., USA) which operates in Resonance Enhanced AFM-IR in contact mode. The AFM scanning was performed by Al and Au coated HQ: CSC38/Al BS probes (Micromasch) having a spring constant of 0.03-0.09 N/m with resolution of 512 x 512 pixels and 128 x co-averages at the rate of 0.6 Hz. AFM-IR images were acquired with pulse rate of 169-171 kHz and frequency window of 79 kHz with resonance Enhanced PLL mode (Phase-Locked-Loop) were feedback loop traced the contact resonance frequency of cantilever. Each spectrum is collected at the targeted position with the spectral resolution of $2 cm^{-1}$ and 128 x co-averages within the range of 1522-1824 $cm^{-1}$ and duty cycle at 5%.

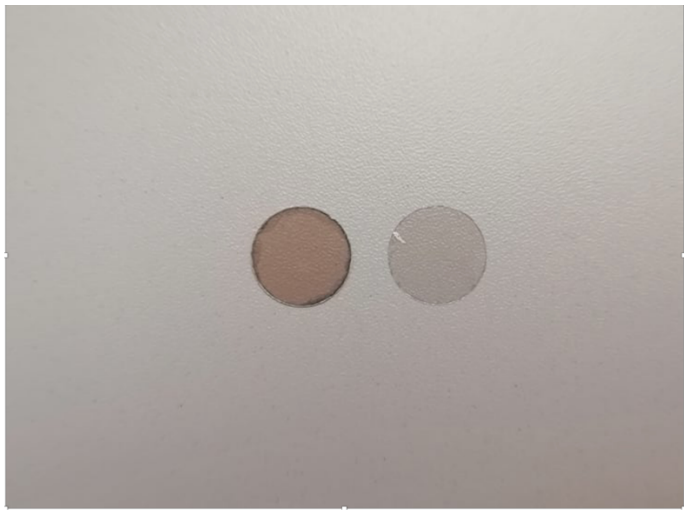

**Figure S4** Photography of muscovite mica substrate: a) Purchased mica substrate (Electron Microscopy Science) with thickness of 0.15-0.21 mm and b) scotch tape exfoliated mica substrate for IR-AFM experimental set-up.

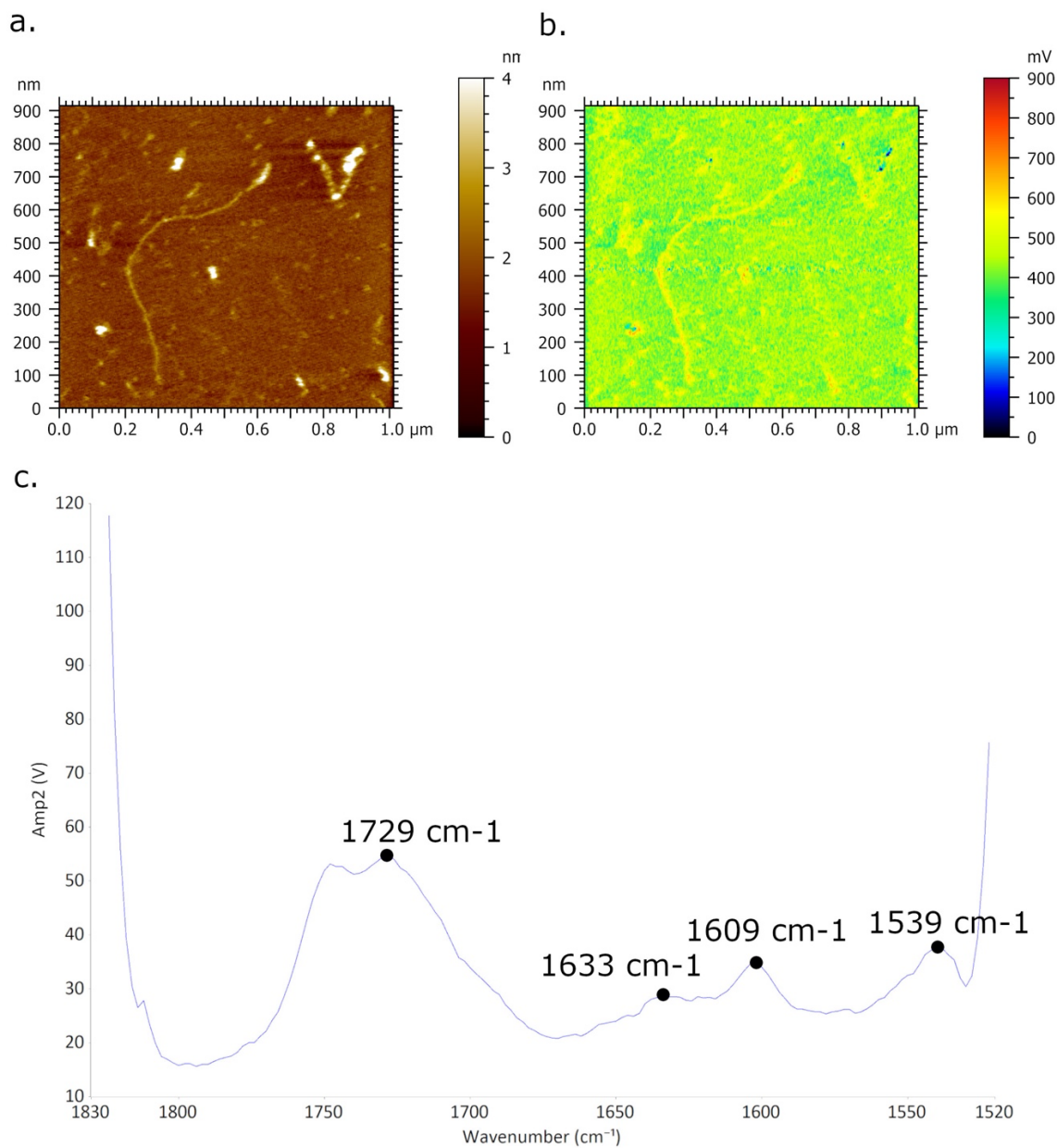

**Figure S5** AFM-IR imaging of single DNA molecule deposited upon Ni$^{2+}$ pretreated mica surface. Size of imaged area: 1.0 x 1.0 μm. a) AFM topography image of DNA and corresponding b) AFM-IR absorption map of single DNA recorded at optimized wavenumber 1728 cm$^{-1}$ showing DNA of higher absorbance (bright red contour) regarding to mica surface. c) AFM-IR spectra acquired and collected from two positions of absorption maxima at 1728 cm$^{-1}$.

As a first step towards acquisition of IR-AFM absorbance map and spectrum of single DNA, we optimized laser focus at 1728 cm$^{-1}$ which correspond to the highest DNA network absorbance among presented series of absorption maps (see Figure 1e). The single DNA bright red contour acquired at 1728 cm$^{-1}$ in figure S5b presents higher DNA absorbance with the respect to the undelaying mica surface. The acquired and averaged spectra (Figure S5a and S5b) based on two positions on DNA molecule (outline by blue crosses in Figure 3a and 3b) obtained broaden spectral band in region between 1672 cm$^{-1}$-1780 cm$^{-1}$ with absorption maxima at 1729 cm$^{-1}$ assigned to C=O stretching[62]. This band, along with shoulder peaks at 1690 cm$^{-1}$ and 1700 cm$^{-1}$ presents overlapping of frequencies assigned base carbonyl stretching and ring breathing mode and double bond in plane stretching involving mainly $C_6=O_6$ carbonyl group of guanines, respectively.

In addition, spectral feature is observed 1633 corresponding to dT, dC, dA residues[63]. Peak with prominent intensity is found at 1603 is assigned to C=N, $NH_2$ adenine stretching while peak with smaller intensity 1571 cm$^{-1}$ is contributed to C=N stretching of adenine[64].

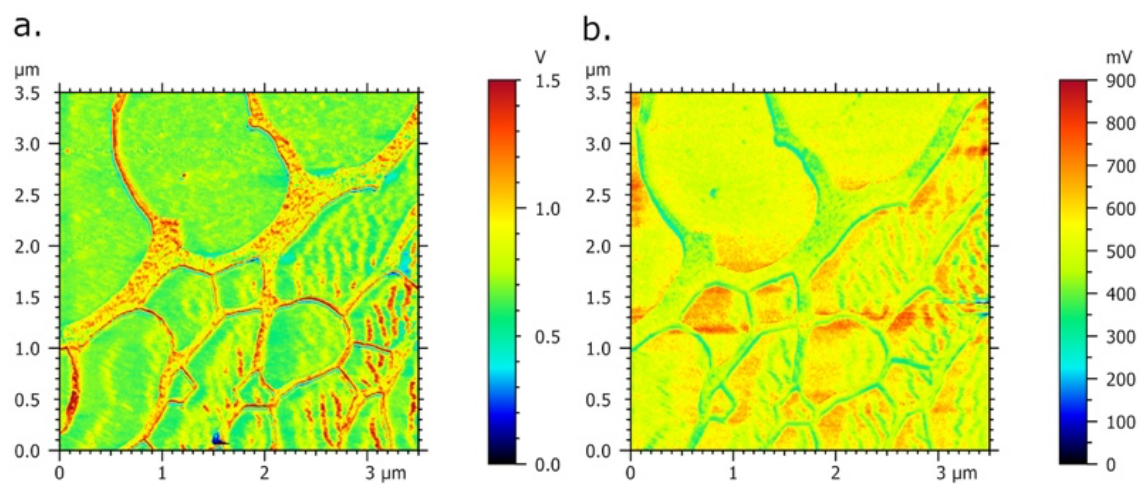

**Figure S6** The effect of ambient humidity on IR-AFM absorption maps of DNA networked upon mica surface. a) IR-AFM absorbance map with higher IR absorbance of DNA (bright red contour) regarding to mica surface (underlying green colorization) mapped at 1728 cm-1. b) IR-AFM absorbance map with lower IR absorbance of DNA (green contour) regarding to mica surface (underlying yellow colorization) mapped at 1728 cm-1 at the ambient of humidity higher than 30 %.

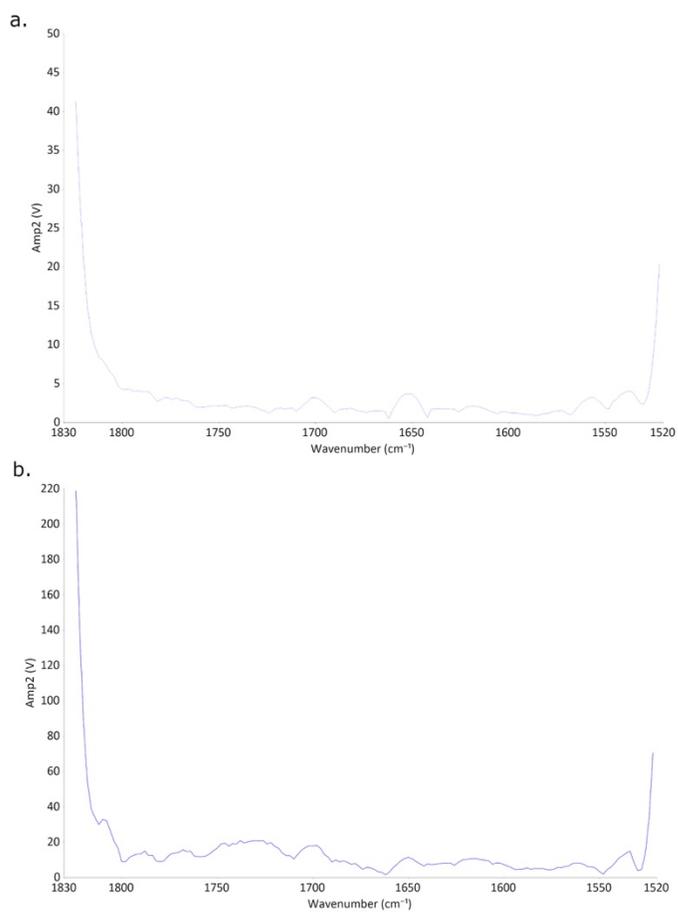

**Figure S7**: IR spectra of a) mica+ spermidine and b) mica + buffers